\documentclass[%
 aip,
 amsmath,amssymb,
 reprint,%
]{revtex4-1}

\usepackage{graphicx}
\usepackage{dcolumn}
\usepackage{bm}
\usepackage{color}
\usepackage[utf8]{inputenc}
\usepackage[T1]{fontenc}
\usepackage{mathptmx}
\usepackage{etoolbox}

\usepackage{graphicx}
\usepackage{epstopdf}
\makeatletter
\def\@email#1#2{%
 \endgroup
 \patchcmd{\titleblock@produce}
  {\frontmatter@RRAPformat}
  {\frontmatter@RRAPformat{\produce@RRAP{*#1\href{mailto:#2}{#2}}}\frontmatter@RRAPformat}
  {}{}
}%
\makeatother

\begin{document}

\preprint{AIP/123-QED}

\title{An efficient method to generate near-ideal hollow beams of different shapes for box potential of quantum gases}
\author{Tongtong Ren}
\affiliation{
Department of Physics and Beijing Key Laboratory of Opto-electronic Functional Materials and Micro-nano Devices, Renmin University of China, Beijing 100872, China}
\affiliation{ 
Key Laboratory of Quantum State Construction and Manipulation (Ministry of Education), Renmin University of China, Beijing 100872, China}
\author{Yirong Wang}
\affiliation{
Department of Physics and Beijing Key Laboratory of Opto-electronic Functional Materials and Micro-nano Devices, Renmin University of China, Beijing 100872, China}
\affiliation{ 
Key Laboratory of Quantum State Construction and Manipulation (Ministry of Education), Renmin University of China, Beijing 100872, China}
\author{Xiaoyu Dai}
\affiliation{
Department of Physics and Beijing Key Laboratory of Opto-electronic Functional Materials and Micro-nano Devices, Renmin University of China, Beijing 100872, China}
\affiliation{ 
Key Laboratory of Quantum State Construction and Manipulation (Ministry of Education), Renmin University of China, Beijing 100872, China}
\author{Xiaoxu Gao}
\affiliation{
Department of Physics and Beijing Key Laboratory of Opto-electronic Functional Materials and Micro-nano Devices, Renmin University of China, Beijing 100872, China}
\affiliation{ 
Key Laboratory of Quantum State Construction and Manipulation (Ministry of Education), Renmin University of China, Beijing 100872, China}
\author{Guangren Sun}
\affiliation{
Department of Physics and Beijing Key Laboratory of Opto-electronic Functional Materials and Micro-nano Devices, Renmin University of China, Beijing 100872, China}
\affiliation{ 
Key Laboratory of Quantum State Construction and Manipulation (Ministry of Education), Renmin University of China, Beijing 100872, China}
\author{Xue Zhao}
\affiliation{
Department of Physics and Beijing Key Laboratory of Opto-electronic Functional Materials and Micro-nano Devices, Renmin University of China, Beijing 100872, China}
\affiliation{ 
Key Laboratory of Quantum State Construction and Manipulation (Ministry of Education), Renmin University of China, Beijing 100872, China}

\author{Kuiyi Gao}%
\altaffiliation[Authors to whom correspondence should be addressed: ]{kgao@ruc.edu.cn}
\affiliation{
Department of Physics and Beijing Key Laboratory of Opto-electronic Functional Materials and Micro-nano Devices, Renmin University of China, Beijing 100872, China}
\affiliation{ 
Key Laboratory of Quantum State Construction and Manipulation (Ministry of Education), Renmin University of China, Beijing 100872, China}

\author{Zhiyue Zheng}
\altaffiliation[Electronic mail: ]{zhengzy@baqis.ac.cn}
\affiliation{ 
Beijing Academy of Quantum Information Sciences, Beijing 100193, China}

\author{Wei Zhang}
\affiliation{
Department of Physics and Beijing Key Laboratory of Opto-electronic Functional Materials and Micro-nano Devices, Renmin University of China, Beijing 100872, China}
\affiliation{ 
Key Laboratory of Quantum State Construction and Manipulation (Ministry of Education), Renmin University of China, Beijing 100872, China}
\affiliation{ 
Beijing Academy of Quantum Information Sciences, Beijing 100193, China}

\date{\today}

\begin{abstract}

Ultracold quantum gases are usually prepared in conservative traps for quantum simulation experiments. The atomic density inhomogeneity, together with the consequent position-dependent energy and time scales of cold atoms in traditional harmonic traps, makes it difficult to manipulate and detect the sample at a better level. These problems are partially solved by optical box traps of blue-detuned hollow beams. However, generating a high-quality hollow beam with high light efficiency for the box trap is challenging. Here, we present a scheme that combines the fixed optics, including axicons and prisms, to pre-shape a Gaussian beam into a hollow beam, with a digital micromirror device (DMD) to improve the quality of the hollow beam further, providing a nearly ideal optical potential of various shapes for preparing highly homogeneous cold atoms. The highest power-law exponent of potential walls can reach a value over 100, and the light efficiency from a Gaussian to a hollow beam is also improved compared to direct optical shaping by a mask or a DMD. Combined with a one-dimensional optical lattice, a nearly ideal two-dimensional uniform quantum gas with different geometrical boundaries can be prepared for exploring quantum many-body physics to an unprecedented level.

\end{abstract}

\maketitle

\section{\label{sec:level1}introduction\protect}

With the continuous development of experimental techniques for trapped ultracold atoms, many exciting progresses have been achieved in the understanding of quantum few-body and many-body physics in different dimensions and configurations~\cite{Dalfovo1999, Giorgini2008, Bloch2008}. However, the conventional harmonic trap leads to an inhomogeneous density distribution, which consequently causes position-dependent energy, length scales, and relative temperature in the trap. This effect might lead to a mixed-phase configuration with atoms in different phases and also brings complications and challenges to the trapping, manipulation, and detection of the system~\cite{Chin2004, Liao2010}. In particular, the features of non-local quantities (such as correlation functions or momentum distributions) are usually smeared out in signals obtained via trap-averaged measurements performed in inhomogeneous traps~\cite{Smith2008}.

These issues encountered in harmonic traps can be naturally overcome by pioneering work of creating homogeneous quantum gases in uniform optical-box potentials~\cite{todd2005, zoran2013}. Many experiments have utilized box traps in various dimensions to produce homogeneous bosonic~\cite{dalibard2015, chin2017, dalibard2018, zorannature2018,  zorannature2021} and fermionic~\cite{MIT2017, henning2018, thomas2019, Li2024} atomic gases to explore new physics, such as dynamical phase transitions, Berezinskii–Kosterlitz–Thouless physics, sound propagation in quantum gases, fermionic pairing, and some other critical problems~\cite{zoran2021}. In these successful attempts, the optical box traps are mainly created by two methods, including hollow beams from fixed optics, e.g., axicons and phase plates, and programmable spatial light modulators such as liquid-crystal modulators (modulate the phase of the light) and digital micromirror devices (modulate the amplitude).

\begin{figure*}
\centering
\includegraphics[width=16.5cm]{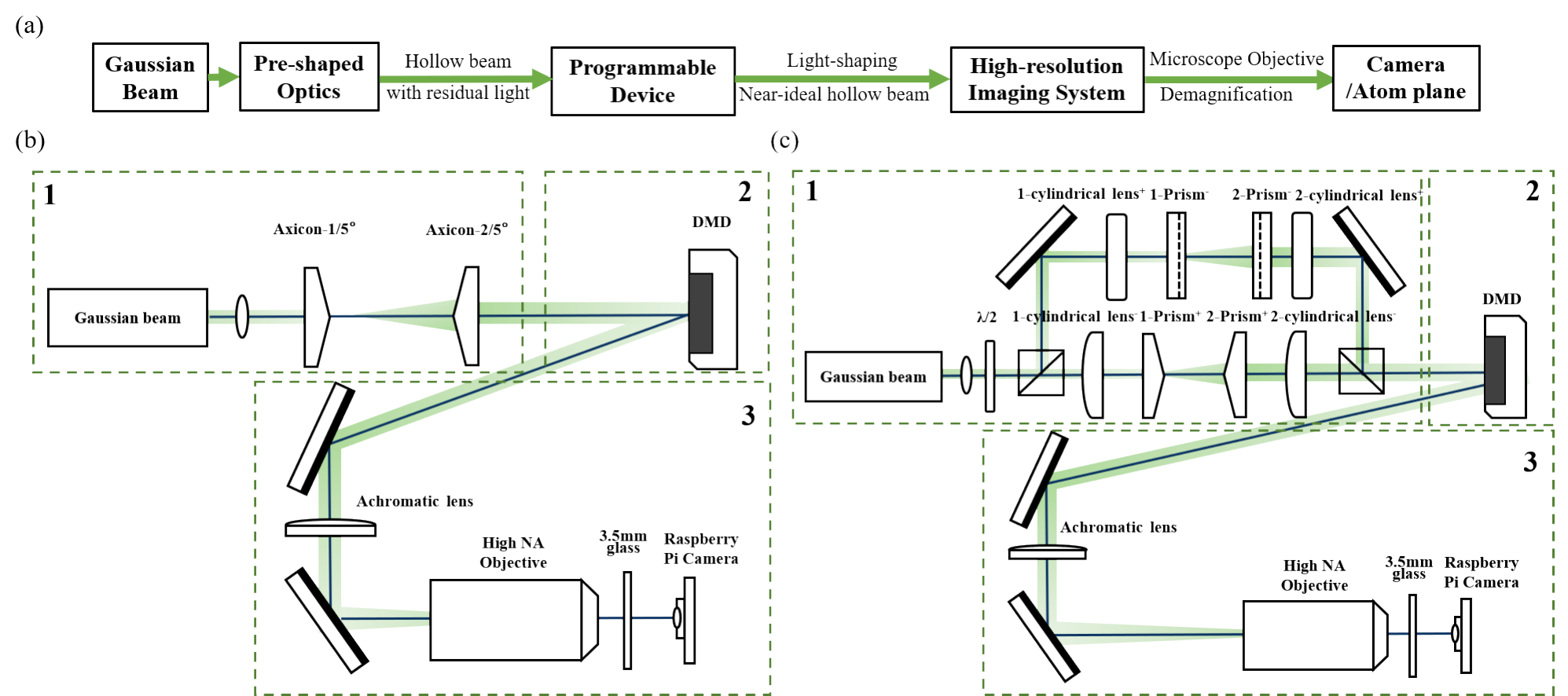}
\caption{\label{fig1} (a) Schematic diagram of the near-ideal hollow beam generation system. (b) The experimental setup of the ring-shaped hollow beam. (c) The experimental setup of the square-shaped hollow beam.}
\end{figure*}

However, these different methods have some drawbacks respectively. The fixed optics approach generates a hollow beam by light deflection or interference, and is often used together with non-adjustable masks of fixed size and shape. The components are usually not programmable, such that it is difficult to get an optimal performance of the resulting box traps. On the other hand, the programmable spatial light modulators, e.g., the DMDs, are easy to adjust by programming. But the conventional schemes usually shape the Gaussian beams directly by cutting a major portion of the beam away. This leads to relatively low optical efficiency, making it difficult to provide high trap depth of far-detuned light for certain experiments due to the limited power of lasers and damage threshold of the DMDs. Moreover, although these methods have been implemented in both two-dimensional and three-dimensional quantum gases in previous experiments~\cite{todd2005, zoran2013, dalibard2015, dalibard2018, chin2017, MIT2017, henning2018, thomas2019, Li2024, zorannature2018, zorannature2021}, the edge of the box potential in three-dimensional situations is far from ideal (represented by a power function with an exponent less than 17)~\cite{zoran2013, MIT2017}, while in two-dimensional cases a much higher value of a ring beam can be obtained with a series of axicons, but a sufficiently good imaging system with an expensive custom made microscopic objective is used~\cite{henning2018}, and this design can only create the ring-shaped box trap.

Here, we present a design with a combination of fixed optics and programmable DMDs to generate near-ideal hollow beams of different shapes for two-dimensional box potential. As shown in FIG.~\ref{fig1}(a), we first use fixed optics such as axicons or multi-facet prisms to get different pre-shaped hollow beams such as a ring or a square. Programmable DMDs are then used to eliminate the stray light in the center of the hollow beam and the residual fringes of the inner side of the ring, so the pre-shaped beam is further improved to a near-ideal hollow shape. Furthermore, a high-resolution imaging system with a commercial high numerical aperture (NA) objective projects the beam to the position of atoms. Such a scheme has many advantages. Most importantly, the hollow beam has a size of tens of micrometers and a steepest beam edge with a power-law exponent of more than 100, which provides a high-quality box for experiments with high precision. The light efficiency from a Gaussian to a hollow beam is increased by a factor of three compared to the direct shaping of the Gaussian beam, which reduces the requirement of the laser power and also lowers the risk of high power damage to the DMD. The programmability of the DMD offers a real-time ability to improve and align the box beam in the experiments. The novel multi-faceted prisms generating hollow beams with different configurations bring more possibilities for future quantum gas experiments and other applications.

\begin{figure*}
\centering
\includegraphics[width=16.5cm]{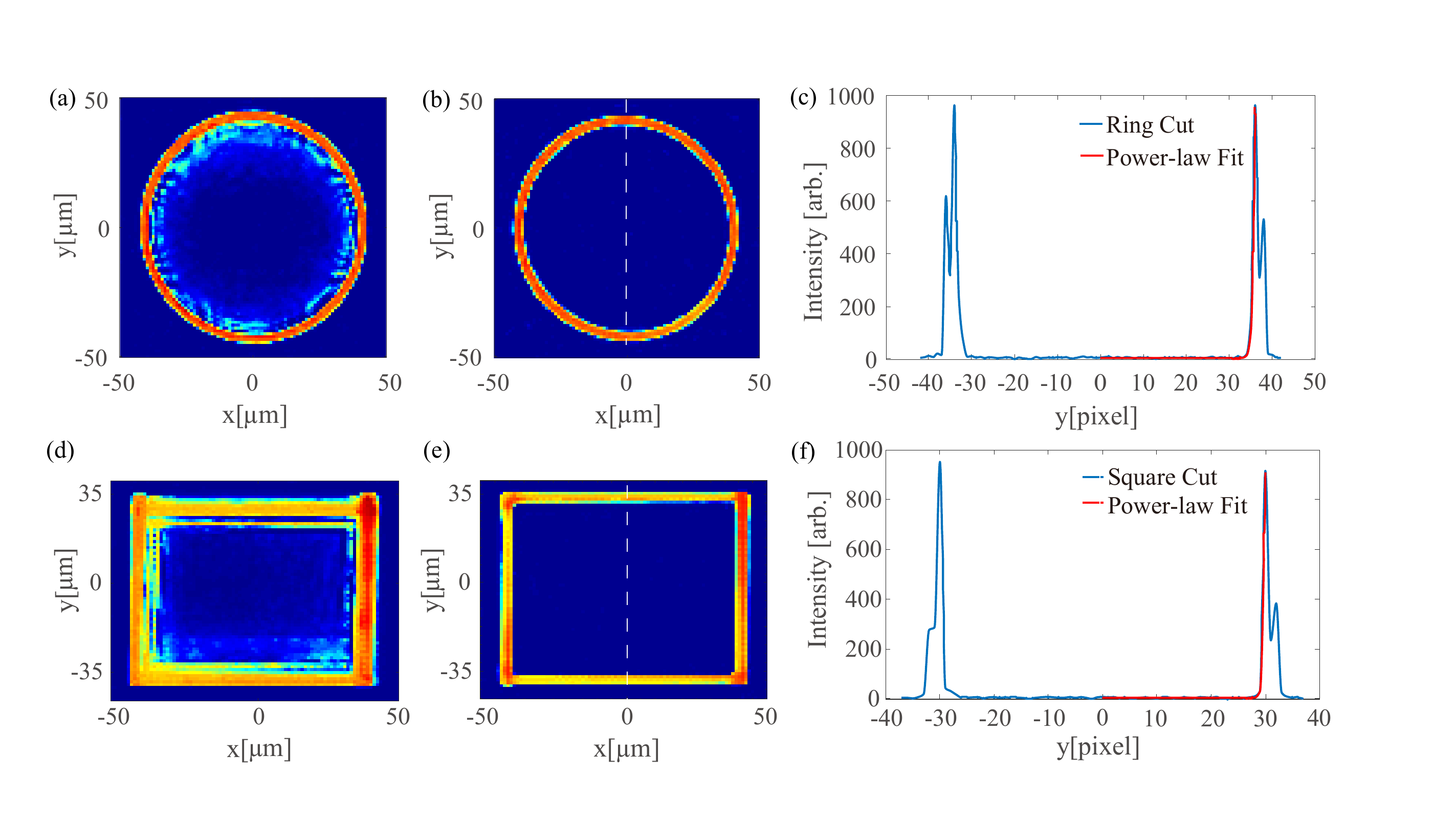}
\caption{\label{fig2} The image of the hollow beam with (b,e) and without (a,d) DMD optimization pictured by the Pi camera at the atom position. For both ring-shaped (a,b) and square-shaped (d,e) cases, the DMD perfectly tailors the residual light inside the hollow beam and realizes a clear and sharp inner boundary. (c) The intensity profile of the ring beam (blue line) along the vertical diameter [white dashed line in (b)]. A power-law fit (red line) gives an exponent of $\alpha = 80 \pm 7$. (f) The intensity profile of the square-shaped beam  (blue line) along a vertical cut through the origin [white dashed line in (e)]. A power-law fit (red) gives an exponent of $\alpha = 104 \pm 9$.}
\end{figure*}

\section{Trap design}

The first stage of our scheme is to use fixed optics, like axicons, to provide the pre-shaped hollow beam. The axicon can reshape the Gaussian beam into a Bessel-like beam in the near field and a ring-shaped beam in the far field~\cite{McLeod1954,axicon1998, MukherjeeThesis, HueckThesis}. Moreover, in the far field, the ring beam propagates under a divergence angle $ \theta\approx2\alpha(n_{g}-1) $, and the ring diameter can be calculated by $ d_{r}=2l \tan[ (n_{g}-1)\alpha ] $ at a distance $ l $ after the tip of an axicon. Here, $\alpha$ and $n_{g}$ are the cone angle and refractive index of the axicon, respectively. Considering the beam width divergence and geometric collimation, in our experiments, a pair of axicons with identical cone angles are positioned on the optical axis with their tips facing each other, leading to a geometrically collimated ring-shaped beam with the minimum divergence of the ring width. By changing the axicons to large-angle multi-faceted prisms, we also demonstrate the generation of hollow beams with different shapes, such as square-shaped beams. Due to the inherent divergence of the Gaussian beam and the inevitable defects at the tip of the axicon from manufacturing, there exist some residual lights in the center and also around the edge of the ring beam, which can significantly degrade the density homogeneity of the trapped atoms. To improve the performance, in the second stage, we use a DMD to remove the residual light and optical fringes inside the ring-shaped beam. Compared to the conventional method using non-adjustable optical masks, the choice of programmable device provides the ability to control the light intensity distribution pixel by pixel. Finally, a telescope with a high NA microscope is used to demagnify the beam from the DMD to the atom position. The high resolution of this imaging system and the high demagnification guarantee the high quality of the hollow beam. 

The optical setup to generate the ring-shaped hollow beam is shown in FIG.~\ref{fig1}(b). The Gaussian beam is from a 532nm laser (Coherent, Verdi V18) and fiber coupled to the test system. The first axicon (Thorlabs, AX255-A) reshapes the Gaussian beam into a ring-shaped beam in the far field. The second axicon collimates the geometry of the ring, by which a pre-shaped hollow ring beam with residual light inside is obtained. The diameter of the ring beam can be adjusted by changing the distance between the two axicons. The pre-shaped ring beam is further reshaped into a high-quality ring-shaped hollow beam by a DMD. In our experiment, we use a DMD (Texas Instruments, DLPLCR67EVM) with each pixel of the chip (DLP670S), having a pixel size of $ 5.4\mu m $, independently set to be ON or OFF states by an external control. The small size of each pixel makes it possible to obtain a sharp inner edge by modifying the pre-shaped hollow beam. The pixels at these two different states reflect the incident beam in different directions. The programmed pattern of the OFF-state pixels acts as a controllable mask for the pre-shaped hollow beam. In experiments, a binary ON-and-OFF image, specifically an ON-state solid circle with the OFF-state background image, is loaded onto the DMD. This ON-state area of the image mostly overlaps with the pre-shaped beam from the fixed optics, and the OFF-state background area just reflects parts of the beam out of the optical axis, which significantly improves the quality of the hollow beam. 

Finally, we rely on a high-resolution imaging system with a plano-convex lens (focal length $ f_{1}=250$mm) and a low cost commercial high NA microscope objective (Mitutoyo, G Plan Apo 50X, NA $=0.5$, $ f_{2}=4$mm) to demagnify the high-quality hollow beam by a factor of $ M={f_{1}}/{f_{2}}=62.5 $ to the atom position. Moreover, a 3.5mm-thick fused silica plate is inserted between the objective and the imaging camera for aberration compensation, which mimics the effect of the glass cell with the same thickness used in the experiment. In order to get a high spatial resolution, a camera with a pixel size of $ 1.12\mu$m (Raspberry Pi, Module V2) is used to measure the light intensity distribution of the hollow beam.

For the square-shaped hollow beam, the overall optical system shares most parts of the one used to generate the ring beam, as shown in FIG.~\ref{fig1}(c). The only difference is that the initial pre-shaped square-shaped beam is generated by combining two pairs of large-angle prisms with their edges placed in horizontal and vertical directions, respectively. A Gaussian beam through a PBS is split into two, and they are turned into two collimated parallel beams with residual light inside by the two pairs of prisms. These two sets of parallel beams are combined into a square-shaped beam through another PBS. The cylindrical lenses after the prisms are used to modify each part of these parallel beams into elongated ones, which makes it easy for subsequent reshaping. After further shaping by the DMD and imaging by the high-resolution imaging system, a nearly ideal square-shaped hollow beam is obtained at the atom position.

\section{Results and Discussion}

The images of the hollow beam at the working distance of the microscope objective are measured by the Pi camera and shown in FIG.~\ref{fig2}. In FIG.~\ref{fig2}(a), the ring beam generated by the axicons is directly imaged without reshaping by the DMD (all DMD pixels are switched to the ON-state and only reflect light). Residual light with diffraction fringes can be clearly observed inside the hollow beam. For the square-shaped beam, a similar phenomenon can also be observed in FIG.~\ref{fig2}(d). By programming the DMD to mask the inner area, a much better pattern can be achieved for both the ring-shaped [FIG.~\ref{fig2}(b)] and square-shaped [FIG.~\ref{fig2}(e)] beams. Both beams present a fairly small size (60-80$\mu m$) and a narrow total width (6-7$\mu m$), with very clean and sharp inner boundaries, as can be observed from the density profile measured along the diameter of the ring [FIG.~\ref{fig2}(c)] and the vertical of the square sides [FIG.~\ref{fig2}(f)]. To further evaluate the performance of our scheme, next we evaluate and analyze some important parameters, including the steepness of the inner edge, the residual light in the center of the beam, and light shaping efficiency, which can all be extracted from the intensity distribution of the hollow beam.

\subsection{Steepness}

One of the most important quality factors of a hollow beam for box potential is the steepness of the inner edge. It can be evaluated by fitting the inner sides of the intensity profile of a cutting-through with a power-law function $ I(x)=Ax^{\alpha} $, with fitting parameters $A$ and $\alpha$. In our experiment, the typical result of the exponent of the power-law fit can reach from $\alpha =81\pm 4$ to $\alpha = 104\pm 10$, indicating the realization of a box potential with a steepest potential $ V(x)=Ax^{104\pm 10} $. The width of the inner edge is carefully measured to be 0.93$\mu m$ by an additional microscope, which is slightly higher than the spatial resolution of 0.67$\mu m$ of our objective, due to the optical aberration of the imaging system. This potential steepness at the edge is higher than that reported in all previous experiments~\cite{zoran2013, MIT2017, henning2018}. Moreover, after subtracting the dark counts of the camera from the beam profile, we find that the average intensity noise from the residual light is less than 2$\%$. Compared to the peak of the hollow beam, it would have a negligible effect on the atomic density in this trap.

\begin{figure}
\centering
\includegraphics[width=9cm]{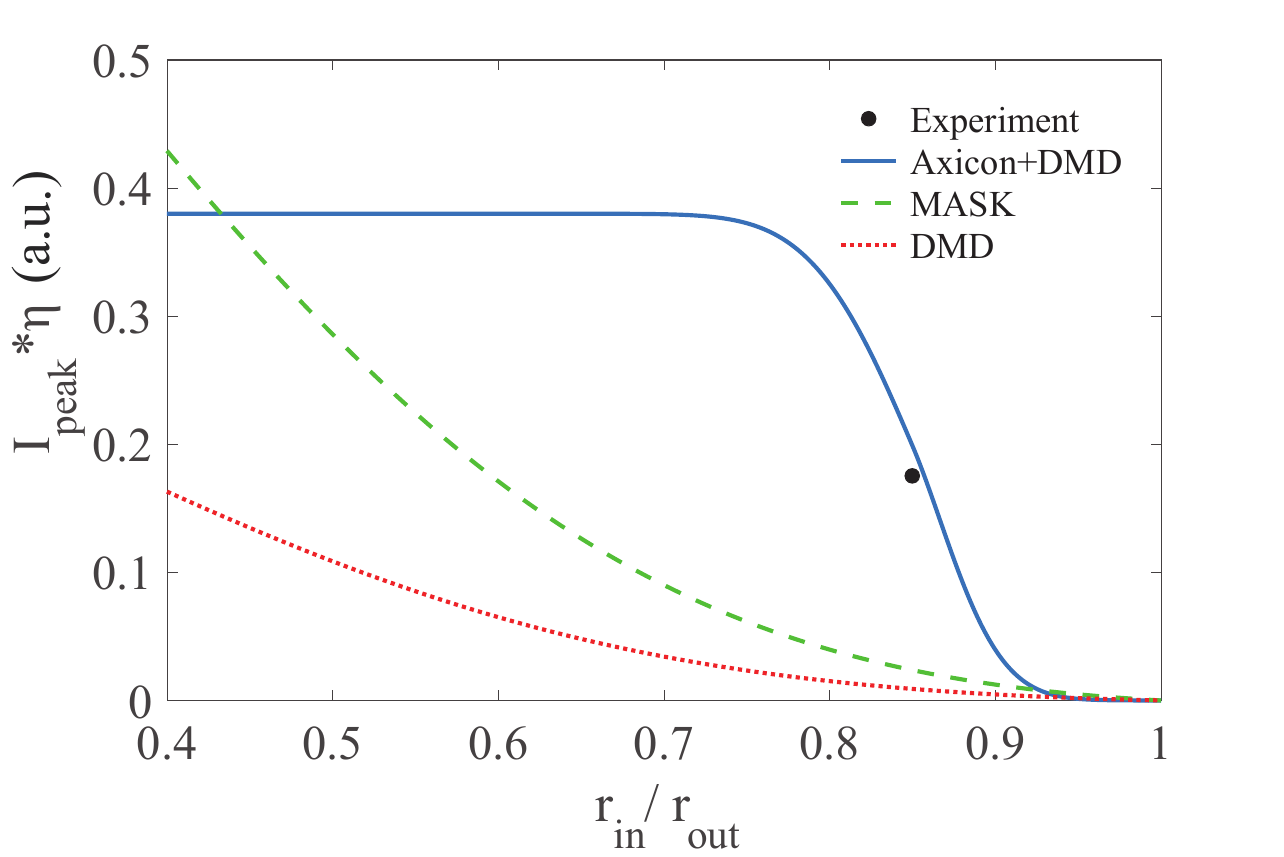}
\caption{\label{fig3} The product of peak-intensity and power efficiency $I_{\rm peak} \cdot \eta$ obtained by different methods, as a function of the ratio between the inner and outer radii of the ring-shaped beam. The blue solid line represents the theoretically calculated result for our proposed scheme, the green dashed line denotes the result from the masking method, and the red dotted line is obtained by using DMD only. The result of DMD is lower than that of the masking method by a factor of 2.7, owing to the diffraction efficiency of DMD of 38\%. Our experimental result (black dot) agrees well with the theoretical prediction and beats the other two methods by a factor of 7.3 at $r_{\rm in}/r_{\rm out}=0.85$.}
\end{figure}

\subsection{Peak Intensity and Efficiency}

Previous experiments that use a mask or a DMD in the direct imaging configuration to generate box traps are essentially performing a direct light-shaping on a Gaussian beam, which deflects most of the light inside the beam out of the optical path. As the center region of the Gaussian beam with most of the power is deflected, this results in a significant loss of optical power, thereby reducing the efficiency of the usually limited laser powers. In our scheme, the Gaussian beam is pre-shaped into a hollow beam by axicons or prisms, where most of the power of the beam is redistributed in the outer region of the beam. Only a small portion of the residual light is left around the inner edge of the pre-shaped hollow beam and is about to be deflected by the DMD. Thus, one would expect a significantly higher efficiency. Moreover, the peak intensity of the hollow beam determines the trap depth of the box potential. For the direct light-shaping of the Gaussian beam by a mask or a DMD, a large cut area in the center of the Gaussian beam leads to a low trap depth of the box potential. However, in our design, the peak intensity of the beam is preserved both in the pre-shaping by the fixed optics and the selective deflecting by the DMD.

To quantify the advantages of our scheme, we define a parameter as the product of peak intensity $I_{\rm peak}$ and efficiency $\eta$. This quantity simultaneously evaluates the performance of the hollow beam and the cost in an energetic aspect, and thus can work as a better assessment of the scheme. In FIG.~\ref{fig3}, we focus on the ring beam as an example and theoretically calculate the product of $I_{\rm peak} \eta$ for different methods, including the direct light-shaping of Gaussian beams by a mask, by a DMD, and our design with a combination of fixed optics and a DMD. Here, the laser efficiency is defined as the ratio between optical power inside the outer radius of the ring and the total laser power $\eta \equiv \int_{ r_{\rm in} < r < r_{\rm out}} I({\bf r}) d{\bf r} / \int I({\bf r}) d{\bf r} $.

For all methods, $I_{\rm peak} \eta$ decreases monotonically with the increase of the ratio between the inner ($r_{\rm in}$) and outer ($r_{\rm out}$) radii of the ring. This behavior is naturally expected as more light needs to be deflected to achieve a ring with a certain size. The most important finding is that for a wide range of $0.43 \lesssim r_{\rm in} / r_{\rm out} \lesssim 0.92$, our method outperforms the other two. Indeed, the $I_{\rm peak}$ remains nearly unchanged for a large parameter region, and starts to drop only at $r_{\rm in} / r_{\rm out} \sim 0.75$ when the DMD begins to cut the inner edge of the pre-shaped hollow beam from the fixed optics. Even for $r_{\rm in} / r_{\rm out} = 0.85$ as demonstrated in our experiment, the new scheme still beats the other two by a factor of at least $\sim 7.3$. The result measured in the experiment (dot in FIG.~\ref{fig3}) agrees very well with the theoretical calculation. For $r_{\rm in} / r_{\rm out} \gtrsim 0.92$, the product $I_{\rm peak} \eta$ approaches zero, since the DMD masks most of the light of the pre-shaped hollow beam such that the efficiency $\eta \to 0$.

\subsection{Shape-variability and Programmability}

The scheme we proposed here combines the advantages from both fixed optics for pre-shaping and DMDs for programmable light-shaping. On one hand, fixed optics can be replaced by elements of different types to generate pre-shaped hollow beams of other geometries, such as pentagon-shaped or hexagon-shaped. Since most of the laser power is already distributed in the hollow beam, the subsequent tailoring by the DMD will not compromise too much on the efficiency. The high efficiency lifts the restrictions of total laser power and damage threshold of optical devices such as the DMD. On the other hand, the usage of DMD offers a great ability to polish the inner edge and form a high-quality hollow beam. Compared to the scheme with a cascade of fixed optics and optical mask to generate hollow beams, the DMD utilized here improves the quality of the hollow beams, without the need to customize different masks for beams with different shapes and sizes. Instead, the programmability of the DMD brings a conveniently adjustable method to further improve the quality of the hollow beam through computer programming. This greatly improves the beam quality while keeping a high light efficiency for light-shaping in the experiments. Furthermore, this programmability also leads to easier alignment of the DMD compared to the mask, where~0.67$\mu$m spatial resolution and stability are needed. With the real-time and ''what you see is what you get" editing ability, our scheme can be integrated with calibration and optimization protocols to build an automated system for generating box potential.

\section{CONCLUSION}
We report a scheme that combines fixed optics and a programmable digital micromirror device to generate high-quality hollow beams of different shapes to realize box potentials for quantum gases. Taking advantage of both fixed optics and programmable DMD, we demonstrate the ability to create a ring-shaped and a square-shaped hollow beam with sharp edges, high power efficiency, and real-time optimization ability. The highest exponent of the light intensity profile of the inner beam edge is 104, outperforming all results reported in previous experiments. Further, we find that the product of peak intensity and laser efficiency is significantly higher than conventional methods in realistic parameter regions, showing the overall performance of our proposed scheme. Our method can greatly improve the uniformity of atomic density in the box potential in the future. With such a high-quality hollow beam and a one-dimensional accordion lattice~\cite{accordion2017}, a homogeneous quasi-two-dimensional box trap for quantum gases might be used to explore its rich physics. This will offer a versatile platform for detailed investigations into quantum phase transitions, coherence, and other many-body effects and advance our understanding of quantum gases and their applications in quantum simulation, quantum computing, and beyond. 

\section*{ACKNOWLEDGMENTS}  
This work is supported by the National Key R\&D Program of China (Grant No. 2022YFA1405301), and the National Natural Science Foundation of China (Grants No. 12274460, 12074428, and 92265208).

~

\section*{DATA AVAILABILITY}  
The data that support the findings of this study are available
from the corresponding author upon reasonable request.

\section*{REFERENCES}
\nocite{*}
\bibliography{idealbox}

\end{document}